\documentclass[12pt,a4paper]{article}
\usepackage{epsfig}
\usepackage{amsmath}

\begin{document}

\title{Cycles structure and local ordering in complex networks} 

\author{Guido Caldarelli$^\star$, Romualdo Pastor-Satorras$^{\dag}$,\\
  and Alessandro  Vespignani$^{\ddag}$ \\ \ \\
  \small $^\star$INFM UdR Roma 1 - Dipartimento di Fisica
  Universit{\`a} "La Sapienza",\\
  \small P.le A. Moro 2, 00185  Roma - Italy\\
  \small $^{\dag}$Departament de F{\'\i}sica i Enginyeria Nuclear\\
  \small Universitat Polit{\`e}cnica de Catalunya, 
  \small Campus Nord,  08034 Barcelona - Spain\\
  \small $^{\ddag}$Laboratoire de Physique Th{\'e}orique (UMR du CNRS 8627)\\
  \small B{\^a}timent 210,  Universit{\'e} de Paris-Sud, 91405 ORSAY Cedex -
  France}

\date{\today}

\maketitle 

\begin{abstract}
  We study the properties of metrics aimed at the characterization of
  grid-like ordering in complex networks. These metrics are based on
  the global and local behavior of cycles of order four, which are the
  minimal structures able to identify rectangular clustering.  The
  analysis of data from real networks reveals the ubiquitous presence
  of a high level of grid-like ordering that is non-trivially
  correlated with the local degree properties.  These observations
  provide new insights on the hierarchical structure of complex
  networks.
\end{abstract}

Many networks arising in social, biological, and technological
contexts are growing and self-organizing systems which are not modeled
by any supervising entity, nor follow an externally defined
blueprint~\cite{barabasi02,dorogorev,watts99,linked,buchanan}.
Empirical evidences, indeed, have prompted that most of the times the
network's topology exhibits complex features which cannot be explained
by merely extrapolating the local properties of their constituents.
The most relevant among these features are the small-world property
\cite{watts98,watts99} and a high level of heterogeneity, usually
reflected in a scale-free behavior of the network's connectivity
\cite{barab99}.  While these properties would prompt to a very large
degree of randomness, yet real networks exhibit a surprising level of
structural ordering. This fact has been first pointed out by noting
the common property of many networks to form cliques in which every
element is linked to every other element; i.e. the presence of a high
clustering coefficient \cite{watts98}, defined as the fraction of
triangles present in the network.  The identification of hidden
ordering and hierarchies in the seemingly haphazard appearance of real
networks is therefore a major area of study, aimed at understanding
their basic organizing principles. This activity has led to a harvest
of results concerning nontrivial correlation properties among the
various elements of natural networks, suggesting the presence of
interesting modular organizations
\cite{alexei,alexei02,assortative,ravasz02}.

In this paper we point out that the usual clustering coefficient is in
some cases unable to quantify the order underlying a network's
structure. In particular, a general modular structure is represented
by a grid-like frame, such as a regular hypercubic lattice, that can
be adequately quantified only by evaluating the frequency of
rectangular loops appearing in the network. We introduce a grid
coefficient that allows us to uncover the presence of a surprising
level of grid ordering in several real networks ranging from
technological (the physical Internet) to social (scientific
collaboration network) systems. By correlating the presence of
grid-like structures with the local connectivity properties we are
able to uncover the presence of a scaling hierarchy that appears to be
a widely present organizing principle. In some cases, the scaling
behavior of the grid clustering is very similar to the triangle
clustering, suggesting a kind of statistical self-similarity in the
modular construction of the network.

A network (or \textit{graph} in the mathematical language
\cite{bollobas98}) is a set of vertices and edges joining pairs of
vertices, representing individuals and the interactions among them,
respectively. Two features play a special role in the characterization
of complex networks. The first one refers to the {\em small-world}
concept \cite{watts98}: i.e. the small average distance in terms of
number of edges between any two vertices in the system. The second
consists in a very high heterogeneity, usually reflected in a
\textit{scale-free} distribution $P(k)\sim k^{-\gamma}$ for the probability
that any given vertex has degree $k$; i.e. $k$ edges to other vertices
\cite{barab99}.  Both properties appear as ubiquitous in dynamically
growing networks \cite{barabasi02,dorogorev}.  Real networks also show
a large degree of local clustering and correlations. A first
quantitative measurements of these properties is provided by the
clustering coefficient \cite{watts98} and the average nearest
neighbors degree \cite{alexei,assortative}. In particular, the
clustering coefficient $c_i$ of the vertex $i$, with degree $k_i$, is
defined as the ratio between the number of edges $e_i$ in the
sub-graph identified by its nearest neighbors and its maximum possible
value, $k_i(k_i-1)/2$, corresponding to a complete sub-graph, i.e.
$c_i=2e_i/k_i(k_i-1)$.  The average clustering coefficient $\langle c \rangle$
is defined as the average value of $c_i$ over all the vertices in the
graph, $\langle c \rangle = \sum_i c_i /N$, where $N$ is the size (total number of
vertices) of the network.  This magnitude quantifies the tendency that
two vertices connected to the same vertex are also connected to each
other; therefore it measures the ordering in the system.  By
comparison, random graphs \cite{erdos59} are not clustered, having $\langle
c \rangle = \langle k \rangle/ N$, where $\langle k \rangle$ is the average degree, while
regular lattices tend to be highly clustered with their neighbors.
Further information can be extracted if one computes the average
clustering coefficient $c(k)$ as a function of the vertex degree $k$
\cite{alexei02}.

In the physics terminology, the study of the clustering coefficient
$c(k)$ is strictly related to the analysis of three-point correlation
functions \cite{alexei03}. The absolute average value---as well as the
scaling with $k$---of this quantity are fundamental to discriminate
the level of randomness and the organizing principles related to the
basic hierarchies present in the networks.  For instance, a large
class of scale-free networks shows a clustering coefficient decaying
as a power-law as a function of the vertex's degree \cite{ravasz02}.
This implies that low degree vertices tend to form connected cliques
with other vertices, while large connected vertices (hubs) tend to act
as bridges between unconnected cliques, thus showing a small
clustering coefficient. This fact highlights the existence of some
modular building, identified by the cliques of small degree vertices
\cite{ravasz02}.

\begin{figure}[p]
  \centerline{\epsfig{file=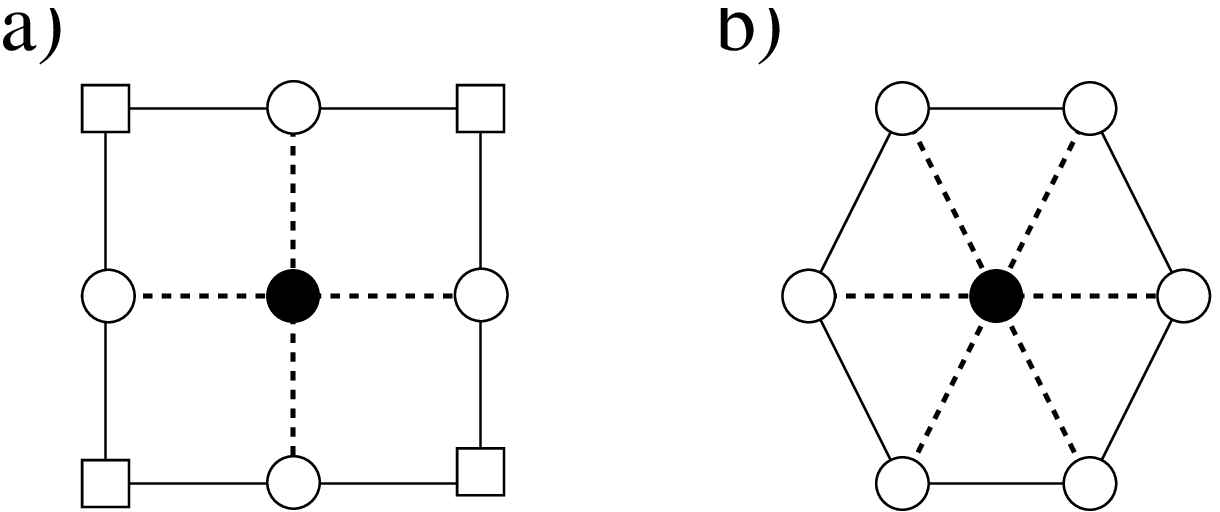,width=10cm}}
  \caption{ \textbf{(a)} Regular square lattice. 
    Nearest neighbors of a vertex (empty circles) are not neighbors of
    each other.  Therefore the clustering coefficient $c_i\equiv 0$ for
    every vertex $i$.  \textbf{(b)} Triangular lattice.  Here some of
    the neighbors are connected to each other. In particular $2$ out
    of every $5$ possible edges are drawn; hence $c_i = 2/5$ for all
    the vertices}
  \label{fig:example}
\end{figure}

\begin{figure}[p] 
  \centerline{\epsfig{file=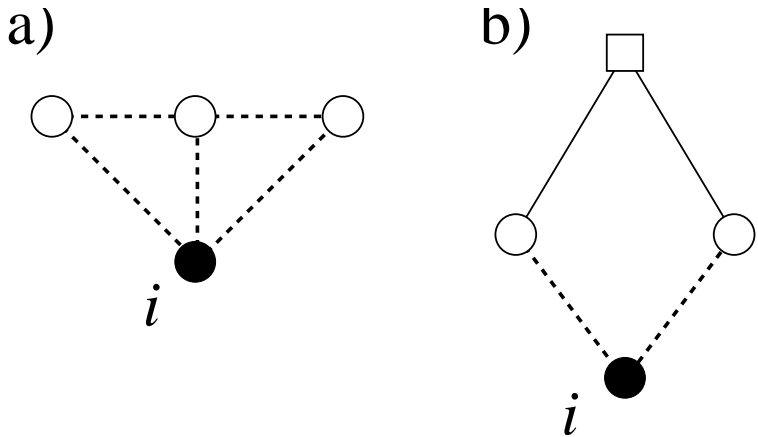,width=10cm}}
  \caption{ \textbf{(a)} Example of a primary quadrilateral, in which
    the three external vertices are nearest neighbors of the vertex
    $i$.  \textbf{(b)} Example of a secondary quadrilateral in which
    one of the external vertices (empty square) is a second neighbor
    of the vertex $i$ .}
  \label{fig:typesofsquares}
\end{figure}

With the aim of unveiling the hidden ordering in complex networks the
use of the two- and three-point correlations is however not always
sufficient. As a very simple example we can consider a rectangular
lattice or grid, Fig.~\ref{fig:example}(a). In this case it easy to
recognize that the clustering coefficient is not able to distinguish
any ordering in a grid-like structure, since its value is always null.
However, it is a good measure of order for other regular structures,
such as a triangular lattice, Fig.~\ref{fig:example}(b).  Since
grid-like structures are among the preferred ordered patterns in
natural systems, we introduce as a further quantitative
characterization of networks' regularity some metrics that naturally
account for rectangular symmetries \cite{vazquez,pusso,biancapo}.  We
start by considering the closed paths in a network in which all edges
and vertices are distinct. These closed paths are known as cycles
\cite{bollobas98}.  Cycles of length $3$ (i.e. composed by three
vertices) are called {\em triangles}. The ratio between the number of
triangles that include the vertex $i$, $e_i$, and its maximum possible
number, $k_i(k_i-1)/2$, defines the triangle coefficient of the vertex
$i$, which is by definition equal to its clustering coefficient $c_i$.
Cycles of length $4$ are called {\em quadrilaterals}.  In the spirit
of the clustering coefficient, we want to improve the measurement of
the network structure by using the \textit{grid coefficient},
$c_{4,i}$, that is defined as the fraction of all the quadrilaterals
passing by the vertex $i$, $Q_i$, divided by the maximum possible
number of quadrilaterals sharing the vertex $i$, $Z_i$. Interestingly,
the grid coefficient can be further decomposed by noting that each
quadrilateral passing by $i$ is composed by the vertex $i$ itself plus
three external vertices.  Quadrilaterals can be therefore classified
according to the nature of the external vertices, see
Fig.~\ref{fig:typesofsquares}. If all the external vertices are
nearest neighbors of $i$, they form a \textit{primary quadrilateral};
on the other hand, if one of the external vertices is a second
neighbor of $i$, the cycle they form is a \textit{secondary
  quadrilateral}. If the vertex $i$ has degree $k_i$ and it is
connected to $k^{nn}_i$ second neighbors, it is easy to check that the
maximum number of primary quadrilaterals is $Z_i^p = 3 \times
\binom{k_i}{3} = k_i(k_i-1)(k_i-2)/2$, while the maximum number of
secondary quadrilaterals is $Z_i^s = k^{nn}_i k_i(k_i-1)/2$.  In this
way, in order to study the grid properties of a network, we can define
three magnitudes: the primary grid coefficient,
$c_{4,i}^p=Q_i^p/Z_i^p$, the secondary grid coefficient
$c_{4,i}^s=Q_i^s/Z_i^s$, and the total grid coefficient $c_{4,i} =
(Q_i^p +Q_i^s)/(Z_i^p+Z_i^s)$, where $Q_i^p$ and $Q_i^s$ are the
actual number of primary and secondary quadrilaterals passing by the
node $i$, respectively. The respective average grid coefficients are
defined by averaging these quantities over all vertices in the
network.

As an example of this definition, let us consider the rectangular
lattice represented in Fig.~\ref{fig:example}(a), in which each vertex
$i$ has $4$ nearest neighbors and $8$ second neighbors.  There are no
primary quadrilaterals passing by any node $i$, while the number of
secondary quadrilaterals is $Q^s= 4$. From here we obtain $\langle c^p_4
\rangle=0$, $\langle c^s_4 \rangle=1/9$, and $\langle c_4 \rangle=1/15$.  On the other hand, in
the triangular lattice, Fig.~\ref{fig:example}(b), in which each
vertex has $6$ nearest neighbors and $12$ second neighbors, we find
$6$ primary quadrilaterals and $6$ secondary quadrilaterals, which
yield $\langle c^p_4 \rangle = 1/10$, $\langle c^s_4 \rangle=1/30$, and $\langle c_4 \rangle=1/20$.
Thus, regular grids exhibit a finite grid coefficient, in opposition
to the clustering coefficient, which is zero for any hypercubic
lattice.

A very different case is represented by the Erd{\"o}s-R{\'e}nyi random graph
\cite{erdos59,bollobas,newmanrev}, constructed from a set on $N$
vertices that are joined in pairs by an edge with probability $p$. In
this case the emerging network has average degree $\langle k \rangle= pN$ and a
Poisson degree distribution, and it is completely random, so that any
ordering is absent in the infinite size limit.  It is possible to
calculate easily the grid coefficients for this network.  For any
vertex $i$, we need at least three nearest neighbors to construct a
primary quadrilateral. Given this configuration, the probability to
close the cycle in any of the three possible quadrilaterals is given
by the probability $p^2$ to draw two edges between two of the three
nearest neighbors. Therefore we have that for any vertex $c^p_{4,i} =
p^2$.  This implies that an Erd{\"o}s-R{\'e}nyi random graph of $N$ vertices
and given average connectivity $\langle k \rangle$ has an average primary grid
coefficient $\langle c^p_4 \rangle =\left( \langle k \rangle / N \right)^2\sim N^{-2}$.  The
calculation for the secondary grid coefficient is slightly more
involved.  In this case, for any vertex $i$, we need at least two
nearest neighbors and a second neighbor. This last vertex, being a
second neighbor, is connected to at least one nearest neighbor, but
not necessarily to any of the two nearest neighbors that will compose
the quadrilateral.  If the second neighbor is not \textit{a priori}
connected to the two nearest neighbors, then the probability of
finding a quadrilateral is of order $p^2$. On the other hand, if it is
\textit{a priori} connected to one of the selected nearest neighbors,
the probability of closing a quadrilateral is just $p$; i.e. the
probability of drawing an edge between the second neighbor and the
remaining nearest neighbor. This last instance (that the second
neighbors is \textit{a priori} connected to one of the nearest
neighbors considered) happens with probability $1/k_i$, where $k_i$ is
the degree of the vertex $i$. Therefore, at leading order in $p$, we
have that for any vertex $c^s_{4,i} = p/k_i$. The average secondary
grid coefficient is then given by $\langle c^s_4 \rangle = \sum_{k\geq 2} P(k) p/k
\equiv \left< 1/k \right>' \langle k \rangle / N$, where $P(k)$ is the degree
distribution of the Erd{\"o}s-R{\'e}nyi random graph \cite{newmanrev}.  By
summing both contributions we have that at leading order, the average
total grid coefficient is scaling as $\langle c_4 \rangle\sim p $. In general thus
the grid coefficient for random graphs of given average connectivity
scales as $\langle c_4 \rangle \sim N^{-1}$ with the number of vertices $N$.

In order to characterize the level of grid-like ordering in real
networks, we have measured the grid coefficients in four different
systems, characterized by a scale-free degree distribution, which have
been the focus of several recent studies:

{\em Internet}: Internet map at the Autonomous System (AS) level, as
of 22nd November 1999 \cite{alexei,alexei02,falou99}.  These maps are
collected and made publicly available by the National Laboratory for
Applied Network Research\footnote{The National Laboratory for Applied
  Network Research (NLANR), sponsored by the National Science
  Foundation, provides Internet routing related information based on
  Border Gateway Protocol data (see http://moat.nlanr.net/).}.  Each
AS refers to one single administrative domain of the Internet.
Different ASs are in most cases connected through a Border Gateway
Protocol (BGP) that identifies any AS through a $16$-bit number.  The
map considered is composed by $6243$ ASs acting as vertices and by
$12 113$ BGP peer connections, acting as edges, yielding an average
degree $\langle k\rangle = 3.88$.

{\em World-Wide-Web}: Map of the World-Wide-Web collected at the
domain of Notre Dame University\footnote{Data publicly available at
  \mbox{http://www.nd.edu/$\sim$networks}.}
\cite{www99,barabasi003,huberman99}. This network is actually
directed, but we have considered it as non-directed. The map is
composed by $325729$ web pages, represented by vertices, and $1090108$
hyperlinks pointing from one page to another, represented by edges,
which corresponds to an average degree $\langle k\rangle = 6.69$.

{\em Movie actor collaborations}: Network of co-actorship obtained
from the Internet Movie Da\-ta\-ba\-se\footnote{The source of the data is
  the Internet Movie Database at \mbox{http://www.imdb.com}. A
  collection of edges where the actors have been numbered is available
  at \mbox{http://www.nd.edu/$\sim$networks}.}
\cite{watts98,barab99,newman01c,amaral}.  In this case the $82583$
actors represent the vertices of the graph. An edge is drawn between two
actors if they have played together in at least one movie. The total
number of edges is $3666738$, with an average degree $\langle k\rangle=88.80$.
 
{\em Scientific collaborations}: Network of scientific collaborations
collected from the condensed matter preprint database at Los
Alamos\footnote{Los Alamos National Laboratory (LANL) preprint
  database located at http://xxx.lanl.gov/archive/cond-mat.}
\cite{newman01a,newman01b,schubert}. This web site hosts the largest
collection of preprints in condensed matter physics. The graph is
composed by $16264$ different authors, that are connected by one edge
if they have coauthored a joint paper.  The total amount of
collaborations (edges) is then $47594$, yielding an average degree $\langle
k\rangle=5.85$.  One can assign a weight to such links, by counting how
many joint paper are present in the repositories. In this first
approach we only considered the topological connections given by the
unweighted links.

In Table~\ref{tab:averages} we report the different average grid
coefficients for all the networks analyzed, compared with those
corresponding to the Erd{\"o}s-R{\'e}nyi random graph, the rectangular grid
and the triangular lattice. It is interesting to note that the average
grid coefficients of all networks are two to four orders of magnitude
larger than the corresponding coefficients of a random graph with the
same average degree and size $N$.  While the small-world property and
the scale-free degree distribution common to all these networks are
generally associated to randomness and large fluctuations, the
presence of large grid coefficient makes those graphs reminiscent of a
grid-like ordering, very probably due to the presence of hierarchical
structures and well-defined communities.

More information can be gathered by studying the grid coefficient as a
function of the vertex's degree $k$ (i.e. by considering the average
value $c_4(k)$ of the total grid coefficient for all the vertices with
the same degree $k$).  As similarly noticed for the clustering
coefficient \cite{alexei02,ravasz02}, the grid coefficient is well
approximated in most cases by a power-law decay for increasing $k$.
This feature indicates a correlation between the vertices' degree and
the local network structure. In particular, low degree vertices are
arranged in fairly ordered patterns whose building blocks are
triangular and rectangular structures.  Vertices with large degree act
as the network backbone by connecting the highly clustered regions.
Since we are facing power-law behavior for the clustering and grid
coefficient, we have that no characteristic length scales are present
in the system and thus there is a continuum hierarchy of structures.
It is also worth remarking that the grid coefficient is always smaller
than the clustering coefficient because it represents a measure of
longer range correlations. Even though statistical fluctuations are
comparable, in some cases the grid coefficient appears to be less
susceptible to noise than other metrics.  Finally, we note the
presence of two classes of networks: the first with a scaling of the
$c_4(k)$ very similar to $c(k)$ (such as the technological Internet
and World-Wide-Web networks), and a second one with $c_4(k)$ different
from $c(k)$ (as is the case of the social networks).  When the
power-law behavior is alike, we can talk of {\em self-similar
  networks} in which both rectangular and triangular patterns are
equally implemented in the modular construction of the network. In the
second situation, one of the two patterns is abandoned earlier in the
hierarchical construction of the graph, breaking the self-similarity
at all levels of the hierarchy.

\section*{Acknowledgments}

The authors wish to thank M. E. J. Newman for making available his
data sets on scientific collaborations. This work has been partially
supported by the European Commission - Fet Open project COSIN
IST-2001-33555. R.P.-S. acknowledges financial support from the
Ministerio de Ciencia y Tecnolog{\'\i}a (Spain).

\newpage

\begin{figure}
\centerline{\epsfig{file=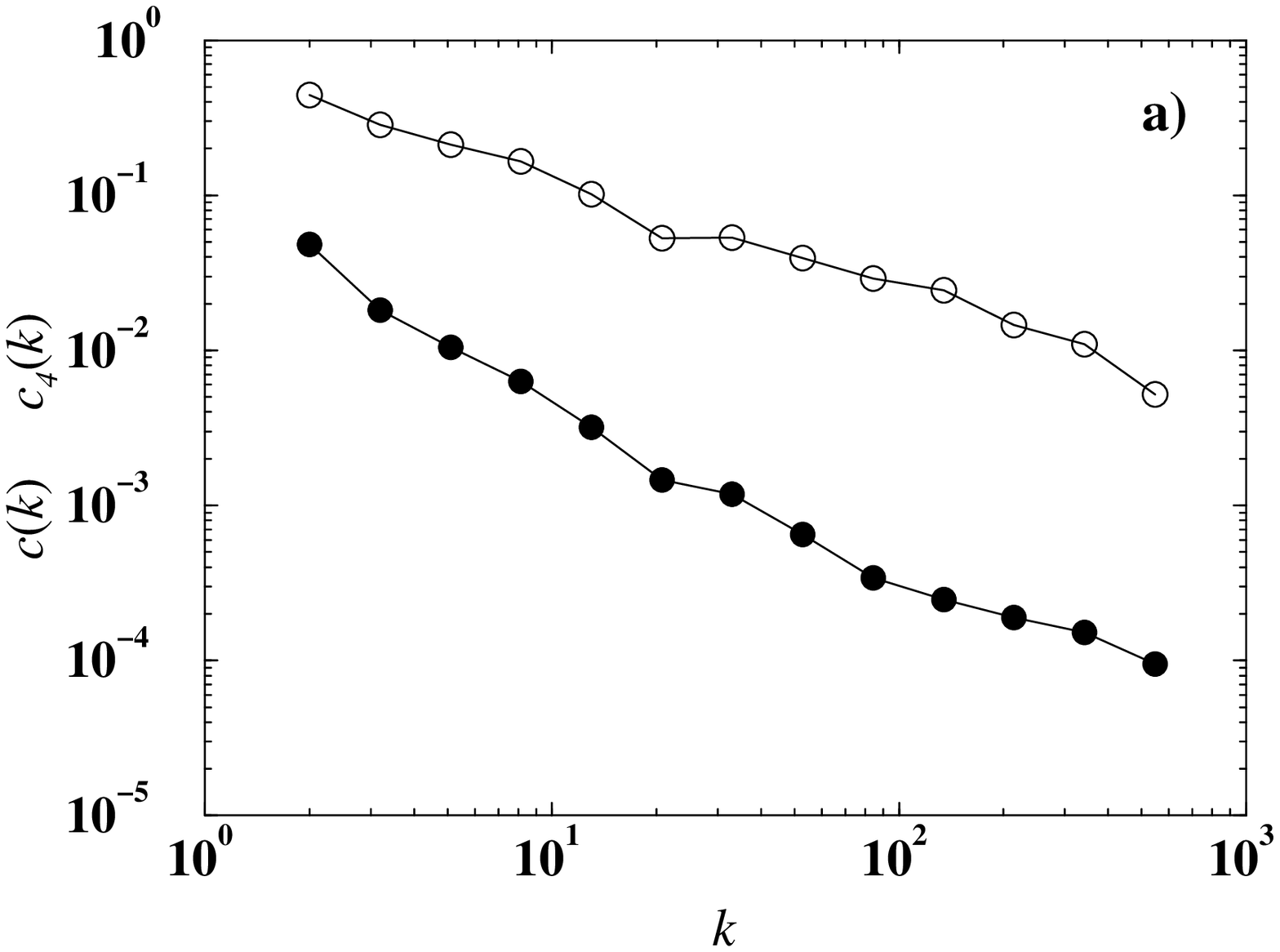,width=7.5cm}%
\hspace*{0.25cm}\epsfig{file=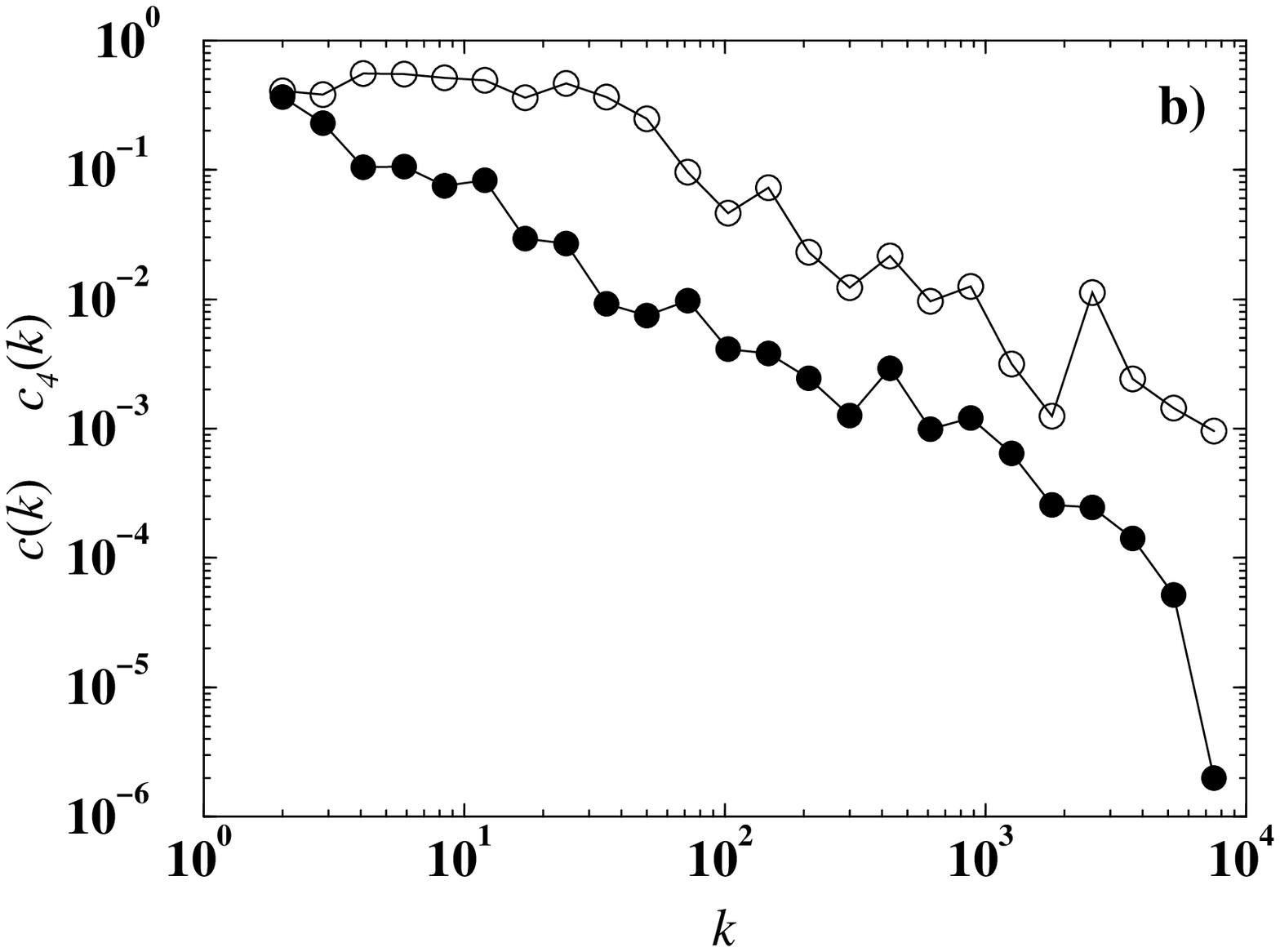,width=7.5cm}}
\centerline{\epsfig{file=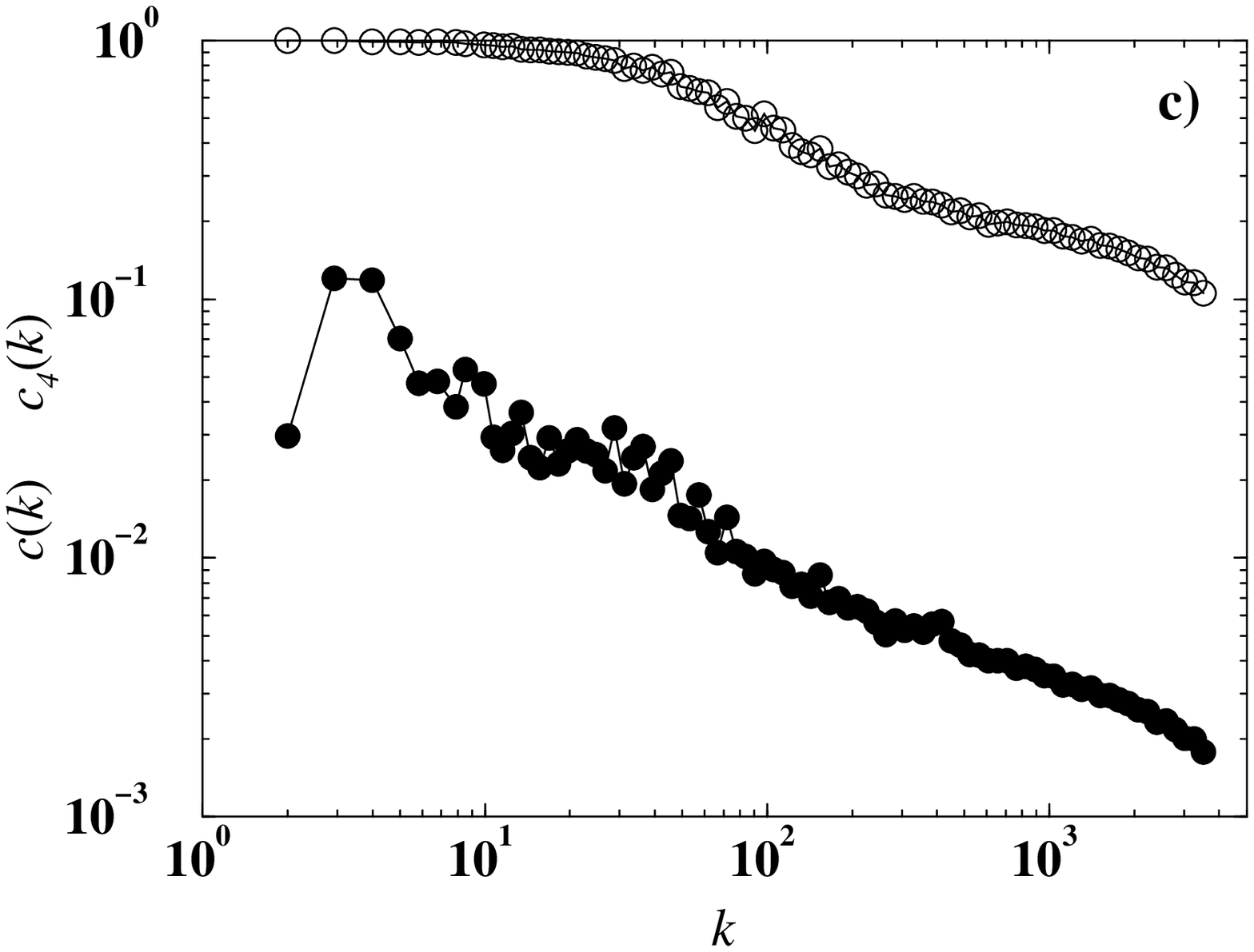,width=7.5cm}%
\hspace*{.25cm}\epsfig{file=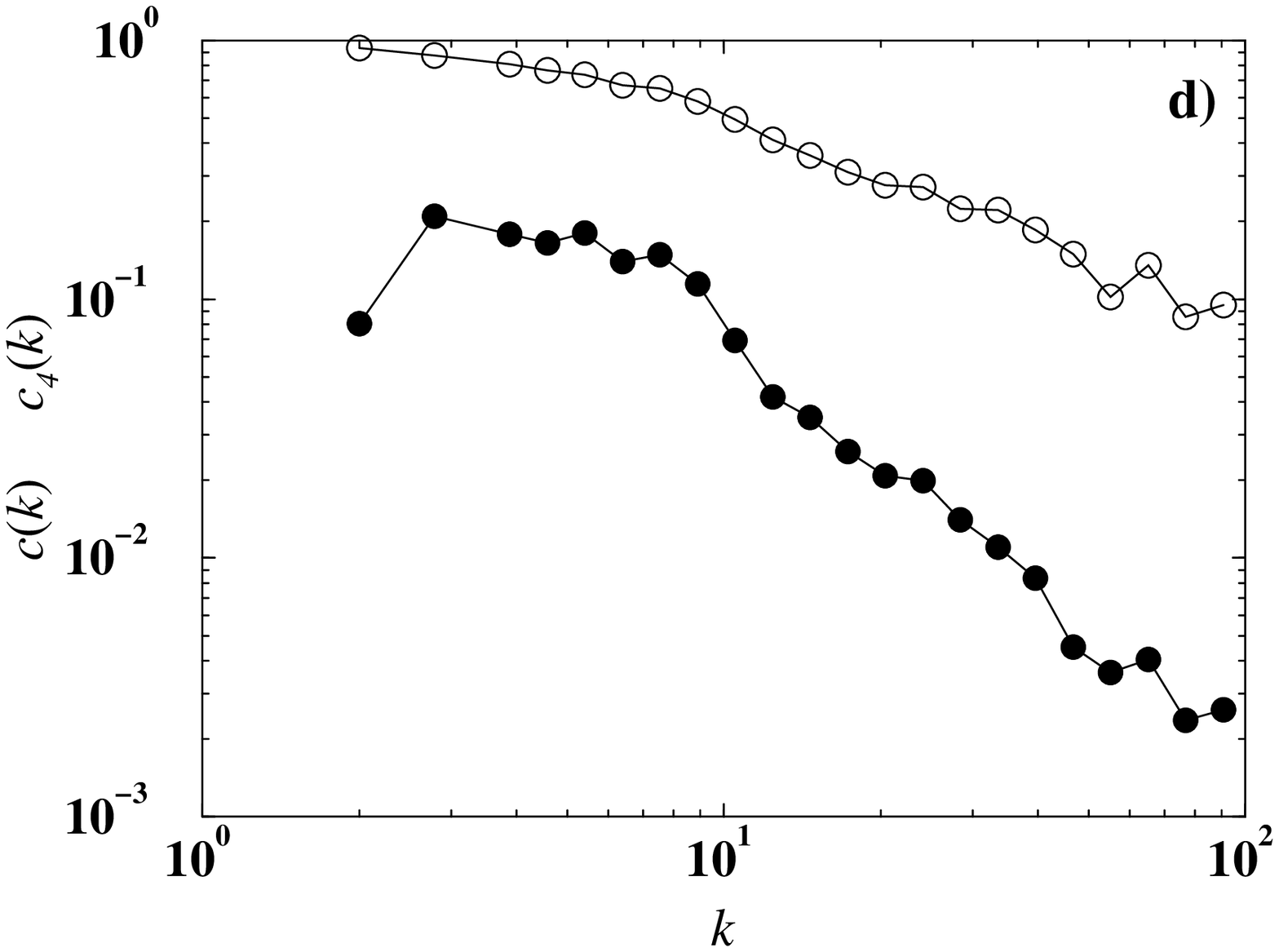,width=7.5cm}}

\caption{Clustering coefficient $c(k)$ (hollow symbols) and grid
  coefficient $c_4(k)$ (filled symbols) as a function of the node
  degree, for the networks considered. \textbf{(a)} Internet at the AS
  level.  \textbf{(b)} Map of the World-Wide-Web domain collected at
  www.nd.edu.  \textbf{(c)} Network of co-actorship from the Internet
  Movie Database. \textbf{(d)} Scientific collaborations from the
  cond-mat preprint database.}
\label{fig:plots}
\end{figure}

\newpage

\begin{table}[p]
\begin{center}
\begin{tabular}{|l|c|c|c|c|c|}
\hline
                   & $\langle k\rangle $ & $\langle
c\rangle$  & $\langle c^p_4\rangle$ & $\langle c^s_4\rangle$ & $\langle
c_4\rangle$ \\ \hline
Internet           & $3.88$  & $0.25$      &  $0.043$ & $0.028$ & $0.028$\\
\hline
World-Wide-Web     & $6.69$  & $0.23$      & $0.14$  & $0.088$ &  $0.090$\\
\hline
movie actor collaborations             & $88.80$ & $0.75$        & $0.66$  & $0.009$ & $0.027$\\
\hline
scientific  coauthorship          & $5.85$  & $0.64$      &  $0.40$ & $0.036$ &
$0.12$\\ \hline  \hline 
Erd{\"o}s-R{\'e}nyi random graph           & $Np$  & $\sim N^{-1}$ & $\sim N^{-2}$  & $\sim
N^{-1}$ & $ \sim N^{-1}$\\ \hline
square lattice     & $4$   & $0$      & $0$  & $1/12$ & $1/15$\\ \hline
triangular lattice & $6$   & $2/5$    & $1/10$  & $1/30$ & $1/20$ \\\hline
\end{tabular}
\end{center}
\caption{Average degree, clustering coefficient, and primary,
  secondary, and total grid coefficients for the different networks
  considered (see text).}
\label{tab:averages}
\end{table}

\end{document}